# Does the AI Productivity Dividend Diminish?

## A Validated Framework for Identifying the Shape of Firm-Level Returns to Artificial Intelligence

Louis Agyekum[1,*]

[1] *Department of Economics, University of Ottawa, Ottawa, ON, K1N 6N5, Canada*

* Correspondence: agyekumlouis9@gmail.com

**Abstract**

Forecasts of the productivity dividend from artificial intelligence (AI) differ by an order of magnitude, partly because they extrapolate average gains observed among early, highly exposed adopters. Whether those gains scale linearly, flatten, or are competed away is an empirical question about the *shape* of the return to AI, not its average. This paper develops and validates a design-based framework for measuring that shape with linked firm–worker data. The framework combines a pre-determined, occupation-based measure of firm exposure to generative AI with a continuous-treatment difference-in-differences design around the release of ChatGPT in November 2022, three pre-specified tests of concavity, a decomposition of exposure effects into an adoption margin and a per-adopter return, a doubly robust analysis of use intensity, a staggered-adoption analysis of earlier AI adopters, and a production-function estimator that places AI in the law of motion of productivity. We validate every component on a calibrated synthetic panel of 20,000 enterprises that reproduces the structure of Statistics Canada's linked business microdata and embeds a known concave effect. The estimators recover the true dose–response curve and detect its concavity (slope difference −0.010, standard error 0.003, one-sided $p = 0.002$); Monte Carlo coverage of the main confidence interval is 0.97, and the concavity test has power of 0.93. The validation also exposes a practical pitfall: pre-trend tests clustered at roughly 40 industries over-reject a true null. The framework, released as open code, is ready for application to Canadian enterprise microdata, where it will inform estimates of potential output and the persistence of Canada's productivity gap.

## 1. Introduction

Artificial intelligence is diffusing through Canadian business faster than almost any general-purpose technology before it. The share of Canadian businesses using AI to produce goods or deliver services rose from 6.1% in the second quarter of 2024 to 12.2% a year later and to 19.2% by the second quarter of 2026, and one in four businesses now plans to use AI within twelve months (Statistics Canada, 2025, 2026). Adoption is concentrated: in 2026 roughly four in ten firms in information and cultural industries and in finance and insurance reported using AI, against a small minority in accommodation and food services.

What this diffusion means for aggregate productivity is far less clear. Acemoglu (2025) combines task-exposure estimates with micro-evidence on cost savings and obtains a total factor productivity (TFP) gain of well under 1% over a decade. Other projections imply gains of more than a percentage point per year. The disagreement is not mainly about whether AI raises productivity where it is used; experimental and quasi-experimental studies show that it often does (Noy & Zhang, 2023; Dell'Acqua et al., 2023; Brynjolfsson et al., 2025). It is about how far those gains travel. Large aggregate estimates implicitly assume that the returns observed among the most exposed early adopters extend, roughly linearly, to firms that are less exposed, adopt later, or use the technology more intensively. If instead the return to AI is concave, so that each additional unit of exposure or use yields less, extrapolation from early adopters overstates the dividend.

This paper is about that shape. We ask whether the productivity return to AI diminishes along three margins: the **intensity** margin (does the effect flatten as firms are more exposed to, or make deeper use of, AI?), the **adoption** margin (do the marginal adopters drawn in by higher exposure gain less than inframarginal ones?), and the **duration** margin (do gains plateau with time since adoption?). Each margin has a distinct economic interpretation. Concavity in intensity points to task saturation or bottlenecks in complementary inputs; declining marginal returns along the adoption margin point to positive selection of early adopters; and plateaus over time distinguish a one-off level shift from sustained growth, a distinction central to the productivity J-curve of Brynjolfsson et al. (2021).

Measuring the shape of a treatment effect is more demanding than measuring its average. Adoption is endogenous, the treatment is continuous rather than binary, there is no untreated control group for generative AI, and firm-level productivity is itself an estimated object. The paper's first contribution is a unified empirical framework that addresses these problems jointly. Its core is a continuous-treatment difference-in-differences design (Callaway et al., 2024) in which treatment intensity is a firm's *pre-determined* exposure to generative AI, built from the payroll-weighted occupational composition of its workforce before November 2022 and published occupation-level exposure scores (Felten et al., 2021; Eloundou et al., 2024). Around this core sit three pre-specified concavity tests, a Wald-type decomposition that separates the adoption gradient from per-adopter returns, a doubly robust analysis of use intensity (Sant'Anna & Zhao, 2020), a staggered-adoption design for an earlier cohort of AI adopters (Callaway & Sant'Anna, 2021), and an Ackerberg et al. (2015) production function in which AI enters the law of motion of productivity (De Loecker, 2013).

The second contribution is validation. Before any confidential data are touched, every estimator is run on a calibrated synthetic panel that mimics the structure of Statistics Canada's National Accounts Longitudinal Microdata File (NALMF), the Canadian Employer–Employee Dynamics Database (CEEDD), the AI modules of the Canadian Survey on Business Conditions (CSBC) and the Survey of Innovation and Business Strategy (SIBS). Because the synthetic data embed a known concave effect, endogenous inputs, selection on unobserved productivity and survivor-biased exit, we can check whether each estimator recovers the truth and whether its inference has the advertised size and power. The estimators perform well: the dose–response curve tracks the true curve, the concavity test rejects linearity with a slope difference of −0.010 (standard error 0.003) against a pseudo-true value of −0.014, Monte Carlo coverage of the 95% interval is 0.97, and the production-function estimator recovers labour and capital elasticities of 0.58 and 0.34 against true values of 0.60 and 0.30. The validation also reveals that a natural and widely used pre-trend test, a joint Wald test clustered at the industry level, rejects a true null far too often when there are only about 40 industries; we document a remedy.

The third contribution is practical. The complete pipeline, from public exposure indices to disclosure-ready output tables, is released as open code and is designed to run unchanged inside Statistics Canada's secure research environment once the synthetic data are replaced by the linked microdata. This matters beyond Canada: secure-enclave research is slow and expensive, and errors discovered after access is granted are costly to fix. Validating a design on synthetic data with known truths before confidential access is a transferable practice.

We stress what the paper does and does not show. The results reported in Section 7 are properties of estimators applied to synthetic data with known parameters; they are not estimates of the effect of AI on Canadian firms. Their purpose is to establish that the framework can detect diminishing returns when they exist, and does not manufacture them when it should not. The next stage of the project is applying this to Canadian microdata, with the research design and pre-analysis plan set out in Section 8.

The rest of the paper is organized as follows. Section 2 reviews the related literature. Section 3 sets out a simple conceptual framework and the hypotheses. Section 4 describes the data architecture and exposure measures, and Section 5 the empirical strategy. Section 6 describes the simulation design. Section 7 reports the validation results, Section 8 discusses implications and the application plan, and Section 9 concludes.

## 2. Related Literature

**Macroeconomic estimates.** A first strand estimates aggregate gains by combining the share of tasks exposed to AI with assumed or measured cost savings per task. Acemoglu (2025) applies Hulten's theorem to such inputs and obtains modest TFP gains, emphasizing that easy-to-learn tasks are likely to be automated first, which itself implies diminishing returns as AI moves into harder tasks. Aghion et al. (2019) show in a growth framework that automation can raise growth only temporarily if non-automated tasks become bottlenecks, a Baumol-type mechanism that also predicts concavity. These approaches are informative but rest on micro-parameters, particularly the curvature of gains across tasks and firms, that are rarely estimated directly.

**Task-level and worker-level evidence.** Experimental and quasi-experimental studies document large gains from generative AI in specific tasks: writing (Noy & Zhang, 2023), consulting (Dell'Acqua et al., 2023) and customer support (Brynjolfsson et al., 2025). A consistent finding is heterogeneity: gains are largest for less experienced or lower-skilled workers and can be negative for tasks outside the technology's frontier. This compression of returns across workers is itself a form of diminishing returns to existing skill. Evidence from Danish administrative data, by contrast, finds small effects of chatbot adoption on earnings and hours (Humlum & Vestergaard, 2025), suggesting that task-level gains do not translate one-for-one into labour-market outcomes.

**Firm-level evidence.** Babina et al. (2024) measure AI investment through the AI skills of firms' employees and find that it raises firm growth, mainly through product innovation, with gains concentrated among larger firms. Czarnitzki et al. (2023) find positive productivity associations of AI use among German firms. Acemoglu et al. (2022) use online vacancies to show that AI-exposed establishments restructure hiring, but find no detectable aggregate employment effects. None of these studies estimates the shape of returns as a function of exposure or use intensity, and most rely on binary adoption or investment indicators that cannot distinguish concave from linear returns.

**Exposure measures and design.** Occupation-level exposure indices (Webb, 2020; Felten et al., 2021; Eloundou et al., 2024) have become standard tools for constructing treatment intensity. Combining pre-period exposure shares with a common technology shock gives a shift-share structure whose identifying assumptions are well understood (Goldsmith-Pinkham et al., 2020; Borusyak et al., 2022). Recent econometric work clarifies what difference-in-differences identifies with staggered adoption (Goodman-Bacon, 2021; Callaway & Sant'Anna, 2021) and with continuous treatments (Callaway et al., 2024), how to combine outcome regression and propensity-score weighting (Sant'Anna & Zhao, 2020), and how to conduct inference that is robust to violations of parallel trends (Roth, 2022; Rambachan & Roth, 2023). On the production side, control-function estimators address the endogeneity of inputs (Olley & Pakes, 1996; Levinsohn & Petrin, 2003; Ackerberg et al., 2015; Gandhi et al., 2020), and allowing a firm decision to enter the law of motion of productivity (De Loecker, 2013) offers a natural way to model AI as a productivity shifter.

This paper brings these strands together. Its distinctive features are a focus on curvature as the estimand, the explicit separation of the adoption margin from per-adopter returns, and validation of the full design against known truths before application.

## 3. Conceptual Framework and Hypotheses

### *3.1 Production with AI*

Consider enterprise *f* in industry *j* and year *t*, with value-added production function

$$y_{ft} = \beta_l l_{ft} + \beta_k k_{ft} + \omega_{ft} + \varepsilon_{ft}, \tag{1}$$

where $y$, $l$ and $k$ are log value added, labour and capital, $\omega_{ft}$ is productivity observed by the firm but not the econometrician, and $\varepsilon_{ft}$ is measurement error. Productivity has a baseline component and an AI component:

$$\omega_{ft} = \omega_{ft}^{0} + A_f \cdot \delta(u_f) \cdot h(t - a_f), \tag{2}$$

where $A_f$ indicates adoption, $a_f$ is the adoption year, $h(\cdot)$ is a phase-in function that rises from zero to one, and $\delta(u_f)$ is the long-run productivity gain at use intensity $u_f$. The central object of the paper is the curvature of δ. Diminishing returns in intensity means $\delta'(u) > 0$ and $\delta''(u) < 0$: deeper use helps, but by progressively less.

### *3.2 From exposure to effects*

Neither adoption nor intensity is randomly assigned. What can be treated as plausibly pre-determined is exposure $E_f$, the extent to which a firm's workforce performs tasks that generative AI can do, measured before the technology became available. Exposure raises both the probability of adoption and the intensity of use. The reduced-form (intent-to-treat) effect of exposure therefore combines two margins:

$$ITT(E) = Pr(A = 1 \mid E) \times \mathbb{E}[\delta(u)h \mid A = 1, E] \tag{3}$$

Equation (3) shows why the curvature of ITT($E$) alone is not decisive. ITT can be concave because adoption saturates as exposure rises, because per-adopter returns flatten, or both. The empirical strategy therefore estimates ITT($E$) and then uses survey information on adoption to separate the two components (Section 5.4).

### *3.3 Hypotheses*

We pre-specify four hypotheses, each stated so that the alternative of constant or increasing returns is equally reportable.

**H1 (intensity).** The dose–response of productivity to AI exposure is concave: $\text{ITT}''(E) < 0$.

**H2 (duration).** For adopters, the dynamic effect of AI rises after adoption and then plateaus, possibly after an initial dip consistent with a productivity J-curve.

**H3 (adoption margin).** The productivity gain of marginal adopters, those induced to adopt by higher exposure, declines as exposure rises.

**H4 (mechanisms and spillovers).** Part of the measured firm-level gain reflects labour substitution or business stealing from less exposed rivals, so that value-added and industry-level gains are smaller than labour-productivity gains of adopters.

## 4. Data Architecture and Measurement

### *4.1 Canadian linked microdata*

The framework is designed for Statistics Canada's linked business microdata, accessed through the Canadian Centre for Data Development and Economic Research (CDER). Table 1 summarizes the files. All are linked through the Business Register enterprise identifier inside the secure environment, and only aggregate outputs that pass disclosure control can be released.

**Table 1.** Linked microdata files and their roles

| File | Unit | Role | Key variables |
|---|---|---|---|
| NALMF | Enterprise × year | Productivity outcomes and inputs | Revenue, GIFI expenses, employment, payroll, assets, NAICS, province |
| CEEDD | Worker × job × year | Pre-period workforce composition; mechanisms | Payroll by occupation, hires, separations, earnings |
| CSBC (AI modules) | Enterprise × wave | Adoption and intensity, 2024–2026 | AI use, applications used, training, workflow changes |
| SIBS | Enterprise × cycle | Earlier AI adoption (pre-2022) | Advanced-technology and AI adoption |
| Public indices | Occupation | Exposure scores | AIOE (SOC); GPT exposure (O*NET-SOC) |

*Notes:* NALMF = National Accounts Longitudinal Microdata File; CEEDD = Canadian Employer–Employee Dynamics Database; CSBC = Canadian Survey on Business Conditions; SIBS = Survey of Innovation and Business Strategy; GIFI = General Index of Financial Information. Public indices are from Felten et al. (2021) and Eloundou et al. (2024).

### *4.2 Exposure to generative AI*

Firm exposure is the payroll-weighted average of occupation-level exposure scores, using the firm's occupational mix in 2019–2021, before the release of ChatGPT:

$$E_f = \sum_o s_{fo}^{2019-21} \cdot Exposure_o\,, \;\; s_{fo} = \frac{Payroll_{fo}}{\sum_{o'} Payroll_{fo'}} \tag{4}$$

The baseline score is the GPT-exposure measure β of Eloundou et al. (2024), which assigns full weight to tasks that a large language model can perform directly and half weight to tasks requiring complementary software. The alternative is the AI Occupational Exposure (AIOE) index of Felten et al. (2021). Using the public files, we harmonize both to six-digit US Standard Occupational Classification (SOC) codes; the merged file covers 889 occupations, and the two measures are strongly but not perfectly correlated (ρ = 0.848). Mapping to Canada's National Occupational

Classification (NOC 2021) uses Statistics Canada's NOC–SOC correspondence, averaging across many-to-many matches; employment-weighted mapping is a robustness check. Because occupational shares are measured before the shock and exposure scores reflect technological feasibility rather than firm choices, $E_f$ is plausibly pre-determined with respect to post-2022 productivity shocks, conditional on industry.

### *4.3 Productivity measures*

Labour productivity is log real value added per worker, with value added equal to revenue less intermediate expenses from the GIFI schedules, deflated by industry price indices, and employment measured in average labour units. TFP is the residual from the production function of Section 5.6. Outcomes are winsorized at the 1st and 99th percentiles within industry-year, a choice varied in robustness checks.

### *4.4 Workflow*

Figure 1 shows the workflow. Public inputs and the synthetic validation run in an open environment; the confidential linkage, variable construction and final estimation run inside the secure environment; and the same estimation code is used in both. Every output passes a disclosure check that records the number of contributing firms per cell before vetting.

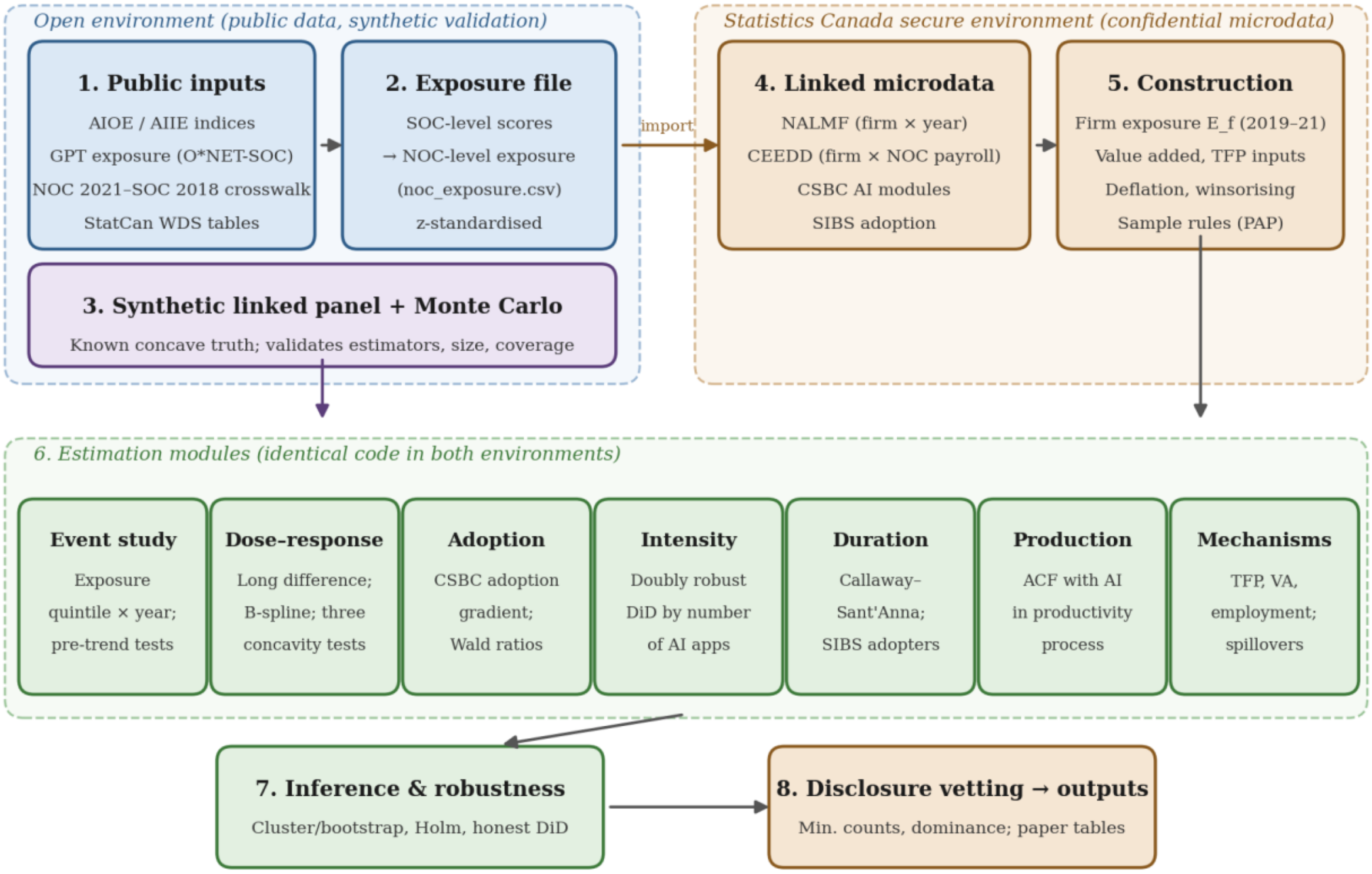

**Figure 1.** Research workflow. Public inputs and synthetic validation (left, blue and purple) run in an open environment; linkage and construction of confidential variables (right, orange) run in Statistics Canada's secure environment. The same estimation modules (green) are applied in both, and all outputs pass disclosure vetting before release.

## 5. Empirical Strategy

### *5.1 Binned event study*

We first divide firms into exposure quintiles $b = 1,\ldots,5$ and estimate

$$y_{ft} = \alpha_f + \lambda_{j(f)t} + \sum_{b=2}^{5} \sum_{k\neq 2022} \beta_{bk} \cdot 1\big[bin_f = b\big] \cdot 1[t = k] + \varepsilon_{ft}, \tag{5}$$

where $\alpha_f$ are enterprise fixed effects and $\lambda_{j(f)t}$ are industry-by-year fixed effects, which absorb industry-specific demand and price shocks. The reference categories are the lowest quintile and 2022, the last full year before generative AI became widely available. Pre-period coefficients $\beta_{bk}$ for $k < 2022$ test parallel trends; post-period coefficients trace the dynamic effect of exposure relative to the least exposed quintile.

Inference requires care. With about 40 industries, a joint Wald test of all 28 pre-period coefficients using industry-clustered standard errors relies on a covariance matrix whose rank is bounded by the number of clusters, and it over-rejects (MacKinnon et al., 2023). We therefore cluster the event study by enterprise and complement the joint test with a low-dimensional test of one linear pre-trend slope per quintile (four restrictions) clustered by industry. Section 7.7 reports the size of both tests.

### *5.2 Dose–response and concavity tests*

The main estimand is the dose–response curve. For each firm we take the long difference between mean outcomes after and before the shock:

$$\Delta y_f = \bar{y}_f^{post} - \bar{y}_f^{pre},\ \ post\ =\ 2023 - 2025,\ pre\ =\ 2019 - 2022, \tag{6}$$

and estimate

$$\Delta y_f = g(E_f) + \mu_j + v_f,\ \ g(E) = \sum_{m=1}^{M} \theta_m B_m(E) \tag{7}$$

where the $B_m$ are cubic B-splines with knots at the quartiles of exposure and $\mu_j$ are industry fixed effects. Under a strong parallel-trends assumption, $g(E) - g(E')$ identifies the difference in average effects between doses $E$ and $E'$ (Callaway et al., 2024). Because every firm has some exposure, levels are not identified without further assumptions, so curves are normalized at the 10th percentile of exposure.

Three concavity tests are pre-specified, each one-sided against the alternative of concavity, using standardized exposure $\tilde{E}_f$. The first is a quadratic test:

$$\Delta y_f = \pi_1 \tilde{E}_f + \pi_2 \tilde{E}_f^2 + \mu_j + v_f,\ \ H_1: \pi_2 < 0 \tag{8}$$

The second, which does not impose a global functional form, is a kinked linear test with the kink at median exposure $m$:

$$\Delta y_f = \kappa_L min(\tilde{E}_f - m, 0) + \kappa_H max(\tilde{E}_f - m, 0) + \mu_j + v_f,\ \ H_1: \kappa_H - \kappa_L < 0 \tag{9}$$

The third replaces exposure with quintile indicators, yielding levels $\theta_b$ relative to the first quintile ($\theta_1 = 0$), and examines their second differences, which are negative under diminishing increments:

$$\Delta_b^2 = \theta_{b+1} - 2\theta_b + \theta_{b-1},\ \ b\ =\ 2, 3, 4 \tag{10}$$

The slope-difference test in equation (9) is the primary test. All three are applied to the stored true effects in the simulations to obtain pseudo-true targets.

### *5.3 Interpretation and identifying assumptions*

The exposure design compares changes in productivity across firms that differed, before November 2022, in how much of their payroll was in occupations that generative AI can perform.

The key assumption is that, absent generative AI, these changes would not have varied systematically with exposure within industries. Its main threats are pre-existing differential trends, for example if exposed firms were already on faster productivity paths, and contemporaneous shocks correlated with exposure, such as post-pandemic shifts in demand for digital services. The event study tests the first directly; industry-by-year fixed effects, the exclusion of the most exposed industries and the alternative exposure index address the second; and sensitivity bounds for violations of parallel trends (Rambachan & Roth, 2023) are part of the pre-analysis plan.

### *5.4 Separating the adoption margin from per-adopter returns*

Equation (3) implies that, if exposure affects productivity only through AI use, the change in ITT between adjacent quintiles divided by the change in adoption rates identifies the average gain of firms induced to adopt by the higher exposure:

$$\rho_b = \frac{ITT_b - ITT_{b-1}}{p_b - p_{b-1}}, \quad b = 2, \ldots, 5, \tag{11}$$

where $p_b$ is the CSBC adoption rate in quintile $b$. A declining sequence of $\rho_b$ is evidence for H3. The ratio is a Wald estimator with a weak first stage wherever $p_b - p_{b-1}$ is small, and its exclusion restriction is strong, since exposure could affect productivity through channels other than AI use. We report it with firm-level bootstrap standard errors that resample the panel and the survey jointly, and only where the adoption gradient is clearly non-zero.

### *5.5 Intensity of use and earlier adopters*

Among CSBC respondents we compare low-intensity adopters (one or two AI applications) and high-intensity adopters (three or more) with non-adopters, using the doubly robust difference-in-differences estimator of Sant'Anna and Zhao (2020):

$$\tau_{DR} = \mathbb{E}\left[\left(\frac{D}{\mathbb{E}[D]} - \frac{\frac{p(X)(1-D)}{1-p(X)}}{\mathbb{E}\left[\frac{p(X)(1-D)}{1-p(X)}\right]}\right)\left(\Delta Y - m_0(X)\right)\right] \tag{12}$$

where $p(X)$ is the propensity score and $m_0(X)$ the outcome regression for non-adopters, with covariates for pre-period productivity level and growth, size, exposure and industry. This comparison relies on selection on observables and is a complement to, not a substitute for, the exposure design.

For firms that adopted advanced AI between 2017 and 2021, observed in SIBS, we estimate group-time average treatment effects with never-treated controls and a universal base period (Callaway

& Sant'Anna, 2021), restricting the sample to 2015–2022 to avoid contamination by generative AI:

$$ATT(g,t) = \mathbb{E}\left[Y_t - Y_{g-1} \mid G = g\right] - \mathbb{E}\left[Y_t - Y_{g-1} \mid C = 1\right],\ \ \theta(e) = \sum_g w_g ATT(g, g+e) \tag{13}$$

with weights $w_g$ proportional to cohort size. A plateau in $\theta(e)$ is the duration version of diminishing returns (H2).

### *5.6 Production function with AI in the productivity process*

To address input endogeneity we estimate equation (1) by the method of Ackerberg et al. (2015), with materials as the proxy variable. The first stage recovers $\varphi_{ft} = y_{ft} - \varepsilon_{ft}$ from a third-order polynomial in labour, capital and materials. The second stage specifies the law of motion of productivity, allowing exposure to shift it after 2022 as in De Loecker (2013):

$$\omega_{ft} = g(\omega_{f,t-1}) + \gamma_1 \tilde{E}_f post_t + \gamma_2 \tilde{E}_f^2 post_t + \psi_j + \xi_{ft} \tag{14}$$

where $g(\cdot)$ is a cubic polynomial and $\psi_j$ captures long-run industry differences. The elasticities solve the moment conditions

$$\mathbb{E}\left[\xi_{ft}(\beta_l, \beta_k) \cdot (k_{ft}, l_{f,t-1})\right] = 0, \tag{15}$$

which hold because capital is chosen before the productivity innovation is realized and lagged labour predates it. A negative $\gamma_2$ is a production-side test of concavity. For markups, the plan applies De Loecker and Warzynski (2012) to a gross-output specification, $\mu_{ft} = \theta^m_{ft} / \alpha^m_{ft}$, where $\theta^m$ is the output elasticity of materials and $\alpha^m$ the materials share of revenue, so that productivity and market-power responses to AI can be separated. Standard errors come from a firm-block bootstrap.

### *5.7 Mechanisms, spillovers and robustness*

The concavity design of Section 5.2 is repeated for TFP, value added and employment to locate where the gains arise. To test for business stealing, low-exposure firms (quintiles 1 and 2) are regressed on the leave-one-out mean exposure of rivals in the same industry and province. A robustness battery re-estimates the primary slope-difference test using the alternative exposure index, the balanced panel, different winsorizing rules, a sample excluding the two most exposed industries and a shorter pre-period, with Holm (1979) adjustment across specifications.

## 6. Simulation Design

The validation uses a synthetic linked panel of 20,000 enterprises in 40 industries and 10 provinces, observed annually from 2015 to 2025. It reproduces the structure of the real files (an enterprise-year panel, an enterprise-occupation payroll table for 2019–2021, two CSBC-like survey waves and a SIBS-like adoption file) and embeds the features that make the real problem hard. Baseline productivity follows a persistent Markov process,

$$\omega_{ft}^{0} = (1-\rho)\mu_f + \rho\omega_{f,t-1}^{0} + \eta_{jt} + \nu_{ft}, \;\; \rho \; = \; 0.85, \tag{16}$$

where $\mu_f$ depends on industry and is correlated with exposure, and $\eta_{jt}$ are industry-year shocks. Capital responds to lagged productivity and labour to current productivity and a persistent adjustment shock, so ordinary least squares production functions are biased. Generative-AI adoption from 2023 depends on exposure and on the firm's unobserved 2022 productivity shock, introducing selection, with a concave adoption gradient:

$$p_f = 0.10 + 0.60\frac{1-e^{-3r_f}}{1-e^{-3}} + 0.08s_f + 0.05z_f, \tag{17}$$

where $r_f$ is the exposure percentile, $s_f$ the standardized 2022 shock and $z_f$ firm size. The productivity gain is concave in use intensity and phases in over three years:

$$\delta(u) = 0.20 \cdot \frac{1-e^{-6u}}{1-e^{-6}}, \;\; h(s) = min\left(\frac{s+1}{3}, 1\right) \tag{18}$$

Adopters also reduce labour slightly. An independent cohort adopts earlier AI between 2017 and 2021, with an effect that dips by 2% in the adoption year and then rises toward a 6% plateau. Firms exit after adverse productivity shocks, producing an unbalanced, survivor-selected panel. For every firm-year, counterfactual outcomes without each AI wave are simulated with the same shocks, so the true treatment effect is known exactly. Table A1 in the Appendix lists all parameters.

Table 2 describes the simulated firms in 2022. Higher-exposure quintiles have higher productivity levels and are larger, mirroring the positive selection that makes naive cross-sectional comparisons misleading; firm fixed effects absorb these level differences.

**Table 2.** Simulated enterprises by exposure quintile, 2022

| Exposure quintile | Firms | Mean exposure *E* | Mean log labour productivity | Mean log employment |
|---|---|---|---|---|
| Q1 (lowest) | 3,209 | 0.367 | 0.181 | 1.394 |
| Q2 | 3,209 | 0.409 | 0.334 | 1.565 |

| Exposure quintile | Firms | Mean exposure *E* | Mean log labour productivity | Mean log employment |
|---|---|---|---|---|
| Q3 | 3,249 | 0.437 | 0.410 | 1.654 |
| Q4 | 3,253 | 0.464 | 0.497 | 1.704 |
| Q5 (highest) | 3,231 | 0.505 | 0.597 | 1.781 |

*Notes:* Synthetic data. The estimation panel contains 182,565 enterprise-years for 17,260 enterprises; 92.3% of enterprises are observed in every year. Of 195 quintile-by-industry cells, 27 fall below the illustrative minimum-count rule, showing why industry detail must be aggregated in released tables.

## 7. Validation Results

All results in this section are obtained from synthetic data with known parameters. They assess whether the estimators recover the truth, not the effect of AI on Canadian firms.

### *7.1 Event study and parallel trends*

Figure 2 plots the event-study coefficients of equation (5) with the true bin-year effects overlaid. Before 2023 the coefficients are small and statistically indistinguishable from zero: the joint test of all 28 pre-period coefficients gives $\chi^2(28) = 23.06$ ($p = 0.730$), and the industry-clustered test of linear pre-trends gives $\chi^2(4) = 1.75$ ($p = 0.782$). After 2022, effects grow in each quintile as adoption and phase-in proceed. Two features of the figure anticipate the main result: the gap between quintiles narrows at the top, and by 2025 the fourth and fifth quintiles are close to each other, as the concave truth implies.

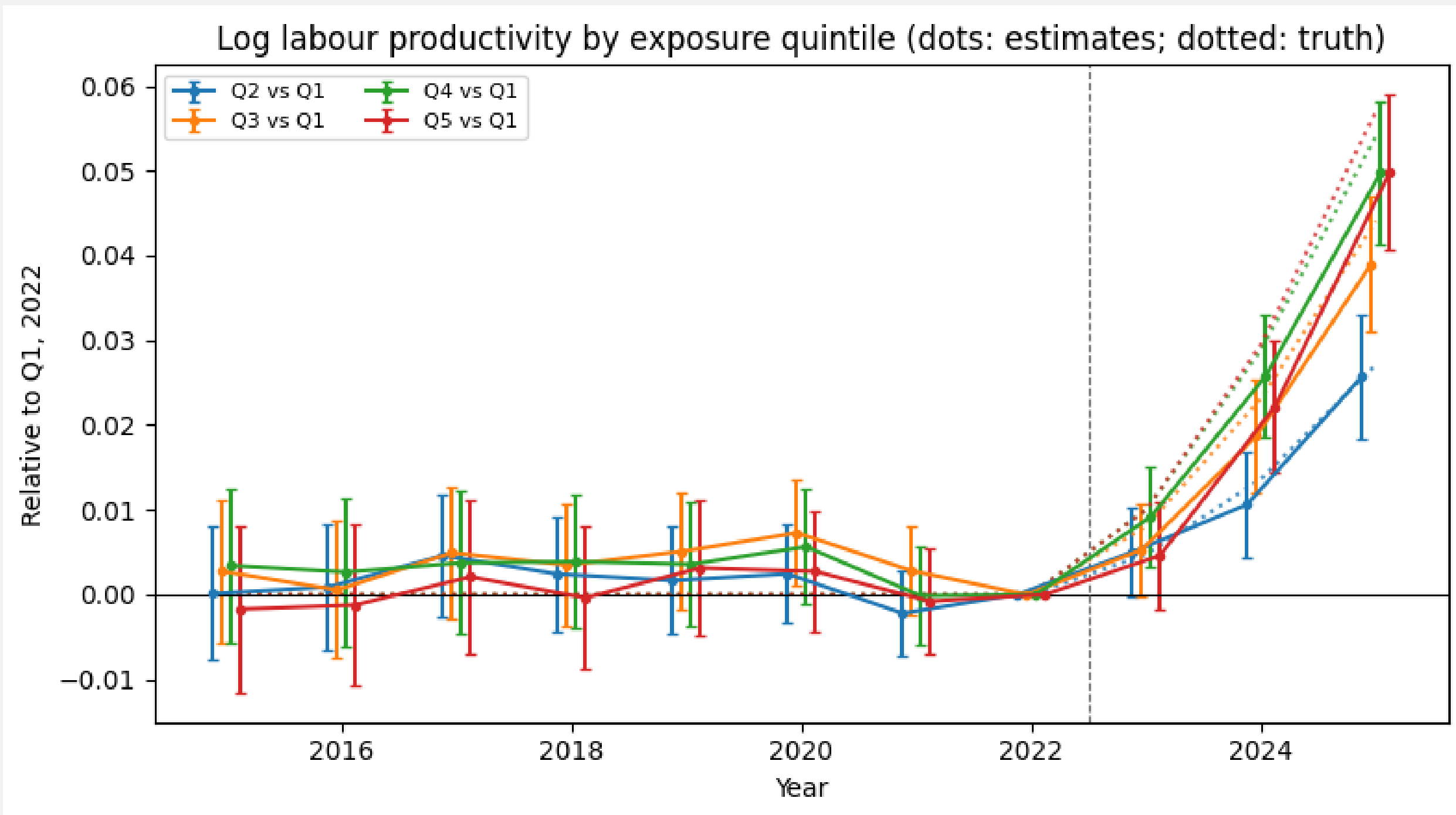

**Figure 2.** Event-study estimates by exposure quintile. Coefficients $\beta_{bk}$ from equation (5), relative to the lowest quintile and 2022, with 95% confidence intervals clustered by enterprise. Dotted lines show the true effects. Synthetic data; 182,565 enterprise-years.

### *7.2 The dose–response curve is concave*

Table 3 reports the concavity tests for 15,823 firms observed in both the pre- and post-periods. All three point in the same direction. The quadratic term is −0.0022 (standard error 0.0009; one-sided $p = 0.007$), against a pseudo-true value of −0.0030. The kinked specification shows the slope falling from 0.0144 below median exposure to 0.0044 above it; the slope difference of −0.0100 (0.0034) rejects linearity ($p = 0.002$). Figure 3 shows the estimated B-spline curve with its confidence band and the true curve. The estimate follows the true shape closely, rising steeply at low exposure and flattening above the median, and the true curve lies within the band over almost the entire support.

**Table 3.** Concavity tests on the dose–response of labour productivity

| **Statistic** | **Estimate** | **Std. error** | **One-sided *p* (concave)** | **Pseudo-true** |
|---|---|---|---|---|
| Quadratic term $\pi_2$ | −0.0022 | 0.0009 | 0.007 | −0.0030 |
| Slope below median $\kappa_L$ | 0.0144 | – | – | 0.0185 |
| Slope above median $\kappa_H$ | 0.0044 | – | – | 0.0043 |
| Slope difference $\kappa_H - \kappa_L$ | −0.0100 | 0.0034 | 0.002 | −0.0143 |

*Notes:* Long-difference regressions (equations 8–9) with industry fixed effects and standard errors clustered by industry; 15,823 firms. Exposure is standardized. Pseudo-true values apply the same specification to the stored true effects.

The pseudo-true values reveal that the estimates are attenuated toward linearity: the estimated slope difference is about 70% of the true one. Two features of the design contribute. Exposure is measured with noise, because the observed occupational mix varies from year to year, and noise in a regressor flattens estimated curvature. In addition, survivor selection removes some low-productivity firms from the long-difference sample. Both biases work *against* finding diminishing returns, so a rejection of linearity in the real data would be conservative.

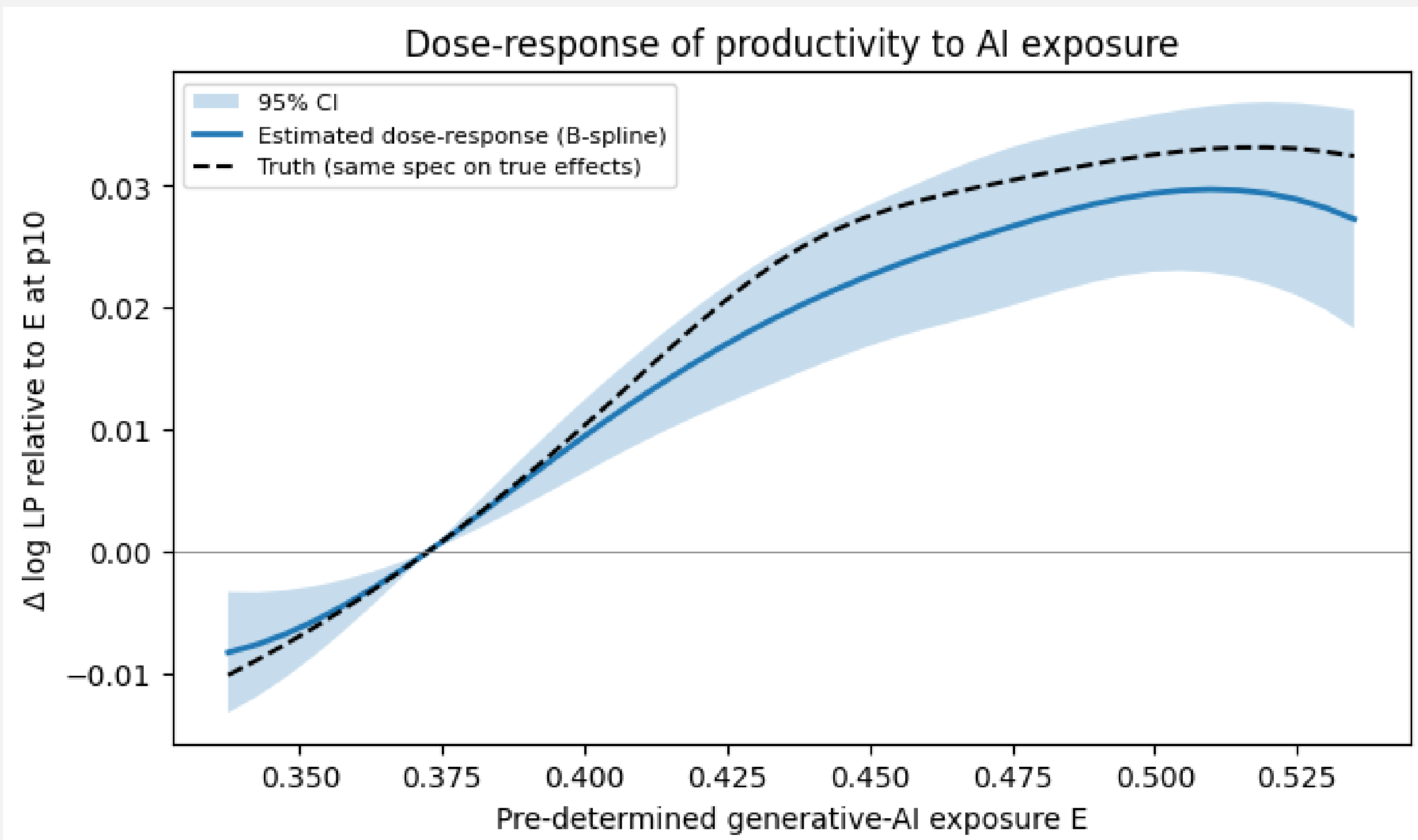

**Figure 3.** Estimated dose–response of labour productivity to AI exposure. Cubic B-spline estimate of g(E) from equation (7), normalized at the 10th percentile of exposure, with pointwise 95% confidence band (industry-clustered). The dashed line applies the same specification to the true effects. Synthetic data; 15,823 firms.

The quintile profile tells the same story, with estimated levels relative to the first quintile of 0.0135, 0.0171, 0.0266 and 0.0247. Second differences are −0.0099 (0.0044), 0.0059 (0.0044) and −0.0114 (0.0046). They are noisier than the parametric tests, as expected from estimating four separate bin effects, which is why the slope-difference test is designated as primary.

### *7.3 The adoption margin*

Table 4 decomposes the exposure effect using adoption rates from the simulated CSBC waves. The adoption gradient is steep at low exposure and flattens at the top, rising from 32% in the first quintile to 49% in the second but only from 70% to 77% between the fourth and fifth. The Wald ratio between the first two quintiles is precisely estimated at 0.077 (0.032). Beyond that, the ratios become imprecise, with bootstrap standard errors above 0.6 in the top two quintiles, because the first stage is weak where adoption saturates. The exercise therefore illustrates both the value and the main limitation of the decomposition: it can isolate the return to marginal adopters only where exposure moves adoption substantially, which argues for coarser bins (terciles) in application.

**Table 4.** Adoption gradient and marginal returns by exposure quintile

| Quintile | Adoption rate $p_b$ | ITT relative to Q1 | Wald ratio $\rho_b$ | Bootstrap s.e. |
|---|---|---|---|---|
| Q1 | 0.317 | 0.0000 | – | – |
| Q2 | 0.492 | 0.0135 | 0.077 | 0.032 |
| Q3 | 0.616 | 0.0171 | 0.029 | 0.090 |
| Q4 | 0.705 | 0.0266 | 0.107 | 0.607 |
| Q5 | 0.770 | 0.0247 | −0.029 | 0.635 |

*Notes:* Adoption rates from the 2025 CSBC-like wave. ITT levels from quintile regressions with industry fixed effects. Wald ratios from equation (11); standard errors from a firm-level bootstrap (199 replications) that resamples the panel and the survey jointly.

### *7.4 Intensity of use*

Table 5 reports the doubly robust comparisons. Both low- and high-intensity adopters show long-difference gains of about 8% relative to non-adopters, close to the true average effects on the treated (0.082 for both groups). The doubly robust estimates (0.083 and 0.081) are slightly closer to the truth than naive difference-in-differences (0.081 and 0.080). In this design, selection on the 2022 productivity shock is modest relative to the effect size, so the correction is small; in real data, where selection is likely to be stronger, the gap between the two estimators is itself informative. That gains are nearly identical for one or two applications and for three or more is consistent with saturation at the intensive margin.

**Table 5.** Doubly robust effects of AI use intensity

| Group | Treated firms | DR ATT | Std. error | Naive DiD | True ATT |
|---|---|---|---|---|---|
| Low intensity (1–2 applications) | 346 | 0.0829 | 0.0052 | 0.0811 | 0.0817 |
| High intensity (3+ applications) | 1,895 | 0.0813 | 0.0033 | 0.0795 | 0.0815 |

*Notes:* Sant'Anna and Zhao (2020) estimator on long differences; comparison group is non-adopting CSBC respondents. Covariates: pre-period productivity level and growth, log employment, exposure and industry indicators. Bootstrap standard errors.

### *7.5 Duration: earlier adopters plateau*

Figure 4 and Table 6 show the Callaway–Sant'Anna event-time effects for the SIBS-era cohort. Pre-adoption coefficients are close to zero. In the adoption year the estimate is −0.015 (0.003), matching the true dip of −0.016, which is the J-curve pattern of costly reorganization before gains appear. The effect then rises to 0.025 after one year and 0.039 after two, with growth slowing thereafter. A quadratic fitted to the post-adoption path has a negative curvature coefficient (−0.0014), consistent with a plateau. Estimates at four and five years are less precise because only the earliest cohorts contribute to them.

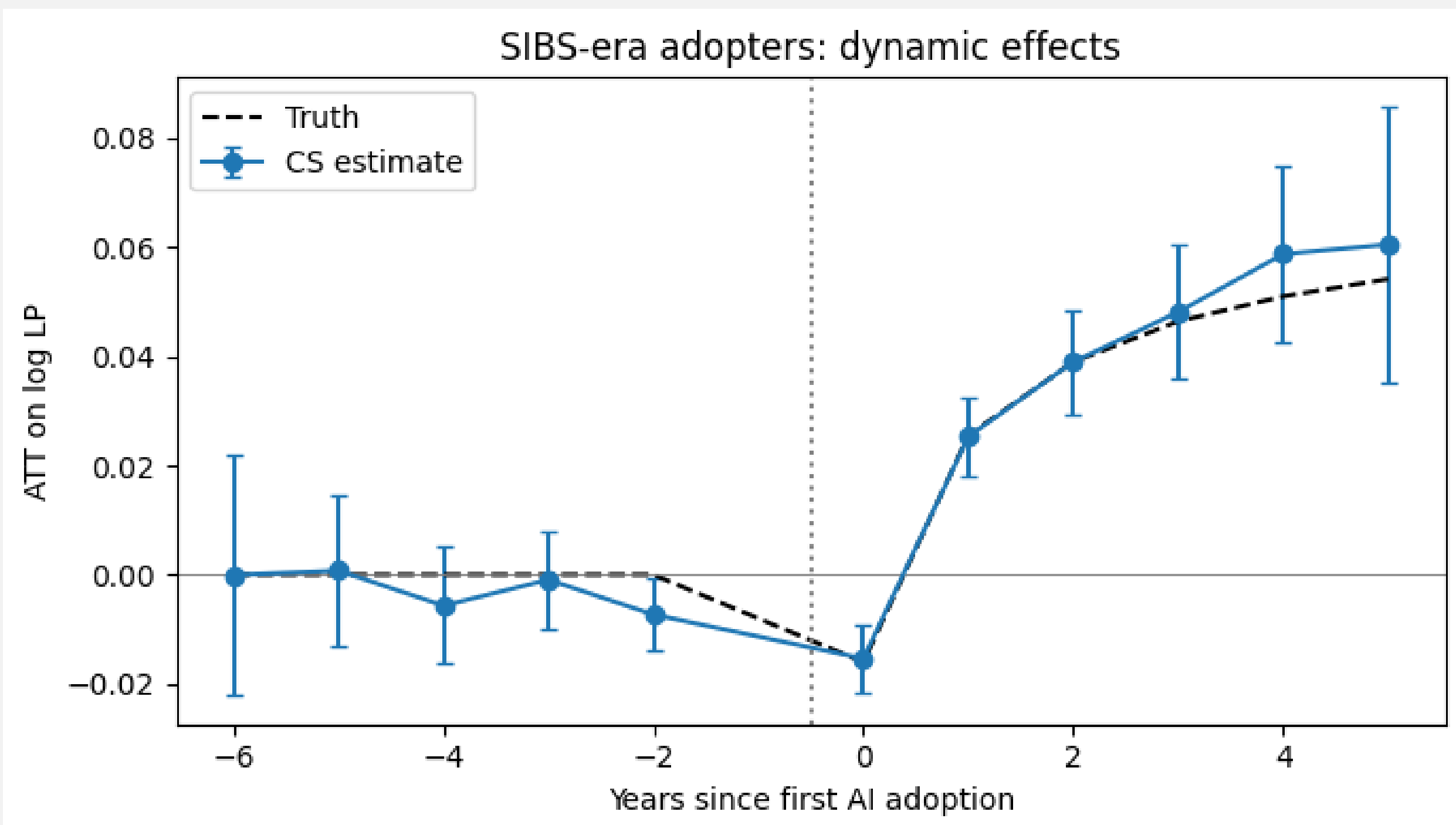

**Figure 4.** Dynamic effects for earlier AI adopters. Callaway and Sant'Anna (2021) event-time aggregates θ(e) from equation (13), never-treated controls, universal base period, 95% bootstrap confidence intervals. The dashed line shows the true effect. Synthetic data, 2015–2022.

**Table 6.** Event-time effects for earlier AI adopters

| **Event time *e*** | **−6** | **−5** | **−4** | **−3** | **−2** | **0** | **1** | **2** | **3** | **4** | **5** |
|---|---|---|---|---|---|---|---|---|---|---|---|
| Estimate | 0.000 | 0.001 | −0.006 | −0.001 | −0.007 | −0.015 | 0.025 | 0.039 | 0.048 | 0.059 | 0.061 |
| Std. error | 0.011 | 0.007 | 0.006 | 0.005 | 0.003 | 0.003 | 0.004 | 0.005 | 0.006 | 0.008 | 0.013 |
| Truth | 0.000 | 0.000 | 0.000 | 0.000 | 0.000 | −0.016 | 0.026 | 0.039 | 0.046 | 0.051 | 0.054 |
| Cohorts | 1 | 2 | 3 | 4 | 5 | 5 | 5 | 4 | 3 | 2 | 1 |

*Notes:* Event time relative to the first year of adoption; $e = -1$ is the base period. Standard errors from a Bayesian bootstrap over firms (199 replications).

### *7.6 Production function and mechanisms*

The Ackerberg–Caves–Frazer estimator with AI in the law of motion recovers a labour elasticity of 0.576 (bootstrap standard error 0.019) and a capital elasticity of 0.337 (0.022), against true values of 0.60 and 0.30. Table 7 applies the concavity design to TFP computed from these estimates, to value added and to employment. Concavity is strong for TFP (slope difference −0.0107, $p < 0.001$) and for value added (−0.0129, $p = 0.001$), showing that the diminishing returns in labour productivity are not an artefact of labour adjustment. Employment effects are small and not significantly concave ($p = 0.210$), consistent with the modest labour substitution built into the design. The rival-exposure coefficient for low-exposure firms is −0.0071 (0.0090), correctly indicating no business stealing in a design that contains none.

**Table 7.** Mechanisms: concavity tests by outcome

| Outcome | Q5 vs Q1 | Slope below median | Slope above median | Slope difference | One-sided *p* |
|---|---|---|---|---|---|
| TFP (ACF) | 0.0246 | 0.0151 | 0.0044 | −0.0107 | 0.0002 |
| Log value added | 0.0279 | 0.0183 | 0.0054 | −0.0129 | 0.0006 |
| Log employment | 0.0026 | 0.0036 | 0.0010 | −0.0025 | 0.2103 |

*Notes:* Long-difference concavity tests (Section 5.2) by outcome, with industry fixed effects and industry-clustered standard errors. TFP uses the elasticities estimated in Section 7.6.

### *7.7 Robustness and Monte Carlo evidence*

Table 8 shows that the primary test is stable across specifications. The slope difference lies between −0.0089 and −0.0116 in every variant, and all remain significant after Holm adjustment (adjusted $p \leq 0.006$). Excluding the two most exposed industries reduces the estimate only slightly, indicating that the result is not driven by a few sectors.

**Table 8.** Robustness of the primary concavity test

| Specification | Firms | Slope difference | Std. error | *p* | Holm *p* |
|---|---|---|---|---|---|
| Baseline | 15,823 | −0.0100 | 0.0034 | 0.0015 | 0.0054 |
| Alternative exposure (AIOE) | 15,823 | −0.0098 | 0.0032 | 0.0010 | 0.0054 |
| Balanced panel only | 15,322 | −0.0104 | 0.0033 | 0.0009 | 0.0054 |
| No winsorizing | 15,823 | −0.0105 | 0.0034 | 0.0011 | 0.0054 |
| Winsorize at 5% | 15,823 | −0.0089 | 0.0031 | 0.0020 | 0.0054 |
| Drop two most-exposed industries | 15,076 | −0.0092 | 0.0034 | 0.0033 | 0.0054 |
| Short pre-window (2021–2022) | 15,823 | −0.0116 | 0.0032 | 0.0001 | 0.0009 |

*Notes:* Each row re-estimates the kinked-linear slope difference (equation 9). One-sided *p*-values; Holm (1979) adjustment across the seven specifications.

Table 9 summarizes the Monte Carlo experiments, which re-simulate the full data-generating process with new seeds. When parallel trends hold, the slope-difference estimator is essentially unbiased relative to its pseudo-true value, its 95% confidence interval covers in 97% of replications, and the concavity test rejects linearity in 93%. Under a differential pre-trend, the joint pre-trend test detects the violation in 70% of replications and the slope test in all of them, demonstrating that the tests have power against the most likely threat.

The experiments also show that the industry-clustered linear pre-trend test rejected a true null in 20% of replications, well above its nominal 5% level, whereas the enterprise-clustered joint test rejected in 10%, within simulation error of 5% given 30 replications. This is a direct illustration of the few-cluster problem (Cameron et al., 2008; MacKinnon et al., 2023): with about 40 clusters, conventional cluster-robust tests can be badly over-sized. In application, pre-trend tests at the industry level should use the wild cluster bootstrap, and the number of replications in this Monte Carlo will be increased to at least 200 for the final appendix.

**Table 9.** Monte Carlo performance of estimators and tests

| Statistic | Parallel trends hold | Differential pre-trend |
|---|---|---|
| Bias of slope difference (estimate − pseudo-true) | 0.000 | 0.002 |
| Root mean squared error | 0.005 | 0.005 |
| Coverage of 95% confidence interval | 0.967 | 1.000 |
| Rejection rate: concavity test | 0.933 | 0.700 |
| Rejection rate: joint pre-trend test (firm clusters) | 0.100 | 0.700 |
| Rejection rate: linear pre-trend test (industry clusters) | 0.200 | 1.000 |

*Notes:* 30 replications with parallel trends and 10 with a differential pre-trend (0.004 per standard deviation of exposure per year), 6,000 enterprises each. Rejection rates at the 5% level.

## 8. Discussion and Application to Canadian Microdata

### *8.1 What the validation establishes*

The simulations establish three things. First, the design can detect diminishing returns when they exist, with high power at sample sizes well below those available in NALMF. Second, it does not manufacture them: under correctly specified dynamics the estimates are unbiased for their pseudo-true values, and the main sources of bias, measurement error in exposure and survivor selection, push toward linearity. Third, inference choices that look innocuous, such as clustering pre-trend tests at the industry level, can be seriously misleading, and the framework documents how to avoid them.

### *8.2 Why the shape matters for policy*

For monetary policy, the shape of the return governs how adoption maps into potential output. If returns are concave, a central bank that extrapolates productivity gains from early, highly exposed adopters to the rest of the economy will overestimate potential growth, understate the output gap and, all else equal, set policy too loosely. The plateau result for earlier adopters matters for the same reason: a level effect that fades into a plateau raises potential output once rather than raising its growth rate permanently, with different implications for the neutral interest rate. For Canada specifically, concave returns would imply that closing the adoption gap with the United States, which Statistics Canada data suggest is now largely closed, will not by itself close the productivity gap; complementary investments in management, data and skills would remain binding.

### *8.3 Pre-analysis plan for the Canadian application*

The application will proceed in the order of Section 7: event study and pre-trend tests, then the primary slope-difference test and dose–response curve, then the adoption decomposition with tercile bins, mechanisms, robustness and the supporting duration and production-function evidence. Sample rules are fixed in advance: employer enterprises outside public administration and non-profit industries, with at least five employees, industry assigned as the modal NAICS code in 2019–2021, and value added and capital deflated with industry price indices. Inference will use at least 999 bootstrap replications, wild cluster bootstrap tests at the industry level, Rambachan and Roth (2023) sensitivity bounds for the event study, and Holm adjustment across outcomes. The hypotheses, specifications and tests in Sections 3 and 5 constitute the pre-analysis plan and will be registered before outcome data are examined.

### *8.4 Limitations*

Several limitations apply. Exposure indices measure technological feasibility, not realized use, and were constructed for the United States; the Canadian occupational mapping introduces additional measurement error. The CSBC is designed as a repeated cross-section, so firm-level adoption histories are partial, and adoption is self-reported. Revenue-based productivity conflates efficiency and prices, which the markup analysis addresses only partly. Tax-based outcomes arrive with a lag, so the initial post-2022 window will be short, which makes the earlier-adopter analysis important for longer horizons. Finally, the synthetic design, however carefully calibrated, cannot anticipate every feature of the real data; its role is to rule out failures of the method, not to predict the answer.

## 9. Conclusion

Whether the productivity dividend from AI diminishes is a question about curvature, and curvature is harder to estimate than an average. This paper has set out an integrated framework for doing so with linked firm–worker data: a pre-determined exposure design around the generative-AI shock, pre-specified concavity tests, a decomposition of exposure effects into adoption and per-adopter margins, and complementary evidence on use intensity, duration and production-side productivity. On a synthetic panel built to share the difficulties of the real data, the framework recovers a concave truth, reports honest uncertainty, and exposes an important pitfall in conventional pre-trend testing. Applied to Statistics Canada's linked microdata, it will deliver direct evidence on whether AI's gains scale with exposure or flatten, evidence that matters for productivity forecasting, for estimates of potential output, and for policy aimed at Canada's persistent productivity gap.

## Declarations

**Data and code availability.** The complete pipeline, including the synthetic data generator, all estimators, Monte Carlo experiments and disclosure checks, is available from the author on request and will be deposited in a public repository upon publication. Public exposure indices are available from their original authors. Statistics Canada microdata are confidential under the *Statistics Act* and are accessible only to approved researchers through Statistics Canada's secure research environments.

**Funding.** This research received no external funding.

**Conflicts of interest.** The author declares no conflicts of interest.

**Use of AI tools.** AI-based assistance was used in drafting code and editing text; the author reviewed all content and takes full responsibility for it.

**Disclaimer.** The views expressed are those of the author and do not necessarily reflect those of any institution with which the author is affiliated.

## References

Acemoglu, D. (2025). The simple macroeconomics of AI. *Economic Policy*, 40(121), 13–58.

Acemoglu, D., Autor, D., Hazell, J., & Restrepo, P. (2022). Artificial intelligence and jobs: Evidence from online vacancies. *Journal of Labor Economics*, 40(S1), S293–S340.

Ackerberg, D. A., Caves, K., & Frazer, G. (2015). Identification properties of recent production function estimators. *Econometrica*, 83(6), 2411–2451.

Aghion, P., Jones, B. F., & Jones, C. I. (2019). Artificial intelligence and economic growth. In A. Agrawal, J. Gans, & A. Goldfarb (Eds.), *The Economics of Artificial Intelligence: An Agenda* (pp. 237–282). University of Chicago Press.

Babina, T., Fedyk, A., He, A., & Hodson, J. (2024). Artificial intelligence, firm growth, and product innovation. *Journal of Financial Economics*, 151, 103745.

Borusyak, K., Hull, P., & Jaravel, X. (2022). Quasi-experimental shift-share research designs. *Review of Economic Studies*, 89(1), 181–213.

Brynjolfsson, E., Li, D., & Raymond, L. (2025). Generative AI at work. *Quarterly Journal of Economics*, 140(2), 889–942.

Brynjolfsson, E., Rock, D., & Syverson, C. (2021). The productivity J-curve: How intangibles complement general purpose technologies. *American Economic Journal: Macroeconomics*, 13(1), 333–372.

Callaway, B., Goodman-Bacon, A., & Sant'Anna, P. H. C. (2024). *Difference-in-differences with a continuous treatment* (NBER Working Paper No. 32117). National Bureau of Economic Research.

Callaway, B., & Sant'Anna, P. H. C. (2021). Difference-in-differences with multiple time periods. *Journal of Econometrics*, 225(2), 200–230.

Cameron, A. C., Gelbach, J. B., & Miller, D. L. (2008). Bootstrap-based improvements for inference with clustered errors. *Review of Economics and Statistics*, 90(3), 414–427.

Czarnitzki, D., Fernández, G. P., & Rammer, C. (2023). Artificial intelligence and firm-level productivity. *Journal of Economic Behavior & Organization*, 211, 188–205.

De Loecker, J. (2013). Detecting learning by exporting. *American Economic Journal: Microeconomics*, 5(3), 1–21.

De Loecker, J., & Warzynski, F. (2012). Markups and firm-level export status. *American Economic Review*, 102(6), 2437–2471.

Dell'Acqua, F., McFowland, E., Mollick, E. R., Lifshitz-Assaf, H., Kellogg, K., Rajendran, S., Krayer, L., Candelon, F., & Lakhani, K. R. (2023). *Navigating the jagged technological frontier: Field experimental evidence of the effects of AI on knowledge worker productivity and quality* (Harvard Business School Working Paper No. 24-013).

Eloundou, T., Manning, S., Mishkin, P., & Rock, D. (2024). GPTs are GPTs: Labor market impact potential of LLMs. *Science*, 384(6702), 1306–1308.

Felten, E., Raj, M., & Seamans, R. (2021). Occupational, industry, and geographic exposure to artificial intelligence: A novel dataset and its potential uses. *Strategic Management Journal*, 42(12), 2195–2217.

Gandhi, A., Navarro, S., & Rivers, D. A. (2020). On the identification of gross output production functions. *Journal of Political Economy*, 128(8), 2973–3016.

Goldsmith-Pinkham, P., Sorkin, I., & Swift, H. (2020). Bartik instruments: What, when, why, and how. *American Economic Review*, 110(8), 2586–2624.

Goodman-Bacon, A. (2021). Difference-in-differences with variation in treatment timing. *Journal of Econometrics*, 225(2), 254–277.

Holm, S. (1979). A simple sequentially rejective multiple test procedure. *Scandinavian Journal of Statistics*, 6(2), 65–70.

Humlum, A., & Vestergaard, E. (2025). *Large language models, small labor market effects* (NBER Working Paper No. 33777). National Bureau of Economic Research.

Levinsohn, J., & Petrin, A. (2003). Estimating production functions using inputs to control for unobservables. *Review of Economic Studies*, 70(2), 317–341.

MacKinnon, J. G., Nielsen, M. Ø., & Webb, M. D. (2023). Cluster-robust inference: A guide to empirical practice. *Journal of Econometrics*, 232(2), 272–299.

Noy, S., & Zhang, W. (2023). Experimental evidence on the productivity effects of generative artificial intelligence. *Science*, 381(6654), 187–192.

Olley, G. S., & Pakes, A. (1996). The dynamics of productivity in the telecommunications equipment industry. *Econometrica*, 64(6), 1263–1297.

Rambachan, A., & Roth, J. (2023). A more credible approach to parallel trends. *Review of Economic Studies*, 90(5), 2555–2591.

Roth, J. (2022). Pretest with caution: Event-study estimates after testing for parallel trends. *American Economic Review: Insights*, 4(3), 305–322.

Sant'Anna, P. H. C., & Zhao, J. (2020). Doubly robust difference-in-differences estimators. *Journal of Econometrics*, 219(1), 101–122.

Statistics Canada. (2025). *Analysis on artificial intelligence use by businesses in Canada, second quarter of 2025* (Catalogue No. 11-621-M). Statistics Canada.

Statistics Canada. (2026). *Analysis on planned use of artificial intelligence by businesses in Canada, third quarter of 2026* (Catalogue No. 11-621-M2026013). Statistics Canada. https://www150.statcan.gc.ca/n1/pub/11-621-m/11-621-m2026013-eng.htm

Webb, M. (2020). *The impact of artificial intelligence on the labor market* (Working paper). Stanford University.

## Appendix A. Simulation Parameters

**Table A1.** Parameters of the synthetic data-generating process

| Component | Specification | Value |
|---|---|---|
| Panel | Enterprises; industries; provinces; years | 20,000; 40; 10; 2015–2025 |
| Occupations | Number; exposure distribution | 40; Beta(2, 2.5) |
| Production | Labour and capital elasticities ($\beta_l$, $\beta_k$) | 0.60; 0.30 |
| Productivity | Persistence ρ; innovation s.d.; industry-year shock s.d. | 0.85; 0.06; 0.03 |
| Inputs | Capital response to lagged productivity; labour shock persistence | 0.25; 0.60 |
| Generative AI | Adoption gradient (eq. 17); adoption years | Concave; 2023–2025 (40/35/25%) |
| Generative AI | Effect scale; curvature; phase-in | 0.20; 6; three years |
| Labour substitution | Share of productivity gain offset in labour | 0.25 |
| Earlier AI (SIBS) | Adoption share; years; dip; plateau | 15%; 2017–2021; −0.02; 0.06 |
| Exit | Logit in productivity shock relative to firm mean | Intercept −4.5; slope −15 |
| Surveys | CSBC waves and sampling rate; SIBS sampling rate | 2024, 2025, 12%; 30% |

*Notes:* Full code is available from the author. Counterfactual outcomes for each AI wave are simulated with identical shocks, so true effects are known for every enterprise-year.